\RequirePackage{fix-cm}
\documentclass[smallextended]{svjour3} 
\smartqed 
\usepackage{graphicx}
\usepackage{tabularx}
\newcolumntype{L}{>{\raggedright\arraybackslash}X}
\usepackage{adjustbox, rotating}
\usepackage{caption}

\newcommand{\athena}{{\it ATHENA }}
\newcommand{\xifu}{{\it X-IFU}}
\newcommand{\wfi}{{\it WFI }}
\newcommand{\chandra}{{\it Chandra }}

\newcommand{\xmm}{{\it XMM-Newton }}

\newcommand{\geotail}{\textit{GEOTAIL}}
\newcommand{\wind}{\textit{WIND}}
\newcommand{\artemis}{\textit{ARTEMIS}}
\newcommand{\stereo}{\textit{STEREO}}
\newcommand{\isee}{\textit{ISEE}}

\newcommand{\ahead}{\textit{AHEAD }}
\newcommand{\arembes}{\textit{AREMBES }}
\newcommand{\exacrad}{\textit{EXACRAD}}

\newcommand{\partflux}{p~cm$^{-2}$~s$^{-1}$~sr$^{-1}$~keV$^{-1}$}

\begin{document}

\title{The particle background of the \xifu~instrument 
}


\author{Simone Lotti  \and
 Teresa Mineo  \and
 Christian Jacquey  \and
 Silvano Molendi  \and
 Matteo D'Andrea  \and
 Claudio Macculi  \and
 Luigi Piro  
}


\institute{S. Lotti \at
  INAF, IAPS, Via fosso del Cavaliere 100, 00133, Roma\\
  Tel.: +39 0649964690\\
  \email{simone.lotti@iaps.inaf.it}  
  \and
  T. Mineo \at
  INAF, IASF Palermo, Via U. La Malfa 153, I90146 Palermo, Italy
  \and
  C. Jacquey \at
  IRAP, CNRS-University of Toulouse, Toulouse, France
  \and
  S. Molendi \at
  INAF, IASF Milano, Via E. Bassini 15, I-20133 Milano, Italy
  \and
  M. D'Andrea \at
    INAF, IAPS, Via fosso del Cavaliere 100, 00133, Roma\\
  University of Roma ”Tor Vergata”, Via della Ricerca Scientifica 1, 00133 Roma, Italy 
  \and C. Macculi, L. Piro \at
  INAF, IAPS, Via fosso del Cavaliere 100, 00133, Roma\\
}

\date{Received: date / Accepted: date}

\maketitle

\begin{abstract}
In this paper we are going to review the latest estimates for the particle background expected on the \xifu~instrument onboard of the \athena mission.
The particle background is induced by two different particle populations: the so called ``soft protons'' and the Cosmic rays. The first component is 
composed of low energy particles ($<100s$ keV) that get funnelled by the mirrors towards the focal plane, losing part of their energy inside the filters and inducing 
background counts inside the instrument sensitivity band. The latter component is induced by high energy particles ($>100$ MeV) that possess enough 
energy to cross the spacecraft and reach the detector from any direction, depositing a small fraction of their energy inside the instrument.
Both these components are estimated using Monte Carlo simulations and the latest results are presented here.

\end{abstract}

\section{Introduction}
\label{intro}
\athena is an observatory, large class, ESA X-Ray mission, whose launch is foreseen in 2028 towards the L2 Sun-Earth Lagrangian point. The mission will
exploit $2~m^{2}$ effective area at 1 keV, 12 m focal length, and two complementary focal plane instruments to achieve its ambitious scientific goals \cite{nandra}.
The first instrument is the Wide Field Imager \cite{rau}(\wfi): a DEPFET detector and a powerful survey instrument that will provide imaging in the 0.1-15 keV band 
over a wide field of view ($40\times40~arcmin^{2}$), simultaneously with spectrally and time-resolved photon counting. The second instrument is the X-ray 
Integral Field Unit \cite{barret}, hereafter \xifu, a cryogenic non-dispersive X-ray spectrometer based on a large array ($\sim 4000$ pixels) of Transition Edge Sensors (TES), 
offering 2.5 eV spectral resolution at 6 keV in the 0.2-12 keV energy band, over a 5 arc minutes diameter field of view. In the following we are going to
estimate the particle background for the \xifu.

Among the scientific goals of the mission there is the observation of faint and/or distant sources (like $z>7$ AGNs), cluster outskirts and the 
Warm Hot Intergalactic Medium (WHIM) and, as a consequence, the instrumental background plays a fundamental role among the mission requirements.
The background for any X-Ray instrument will include three main components:

\begin{itemize}
\item The Cosmic X-Ray Background (CXB), which is composed mainly of photons coming from diffuse or unresolved X-ray sources. This component can be reduced 
with high energy resolution to resolve its line component and with high angular resolution to resolve the individual point sources that generate its high-energy component.
\item The soft protons component, which is induced by low energy particles that get funnelled by the optics toward the focal plane. This component can be reduced with the
use of a magnetic diverter which will deflect the charged particles outside the FoV of the instruments.
\item The high energy component, or Non X-ray Background (NXB), which is generated by high energy particles crossing the spacecraft and the instrument, depositing inside it a fraction of their energy.
These particles create secondaries along their way, which can in turn impact the detector inducing further background counts \cite{lotti2014}. This component is addressed with the presence of an Cryogenic Anti-Coincidence 
detector (CryoAC) placed below the main array, and with the use of a passive shielding to reduce the flux of secondary particles.
\end{itemize}

\noindent In this paper we neglect the photons component, and focus on the background induced by charged particles. In the following (Sect.~\ref{sec:1}) we will estimate the flux of soft protons that will impact the \xifu~instrument, 
while in Sect.~\ref{sec:2} we will deal with the NXB and report the latest estimates for the level of this component.

\section{The soft protons background}
\label{sec:1}

The current generation of X-ray telescopes like \chandra and \xmm has shown that low energy "soft" protons can reduce the available exposure times by up to 50\%, 
introducing in addition a poorly reproducible background component \cite{molendi2000,deluca2004,kuntz2014,leccardi2008}. These soft protons enter the mirrors and are concentrated 
towards the focal plane, losing energy in the instruments filters along their path and finally depositing their remaining energy in the detectors.
Their contribution to the residual background will likely be even higher for \athena with respect 
to previous missions, given the larger collecting area of the mirrors. 

\begin{figure*}
 \includegraphics[width=\textwidth]{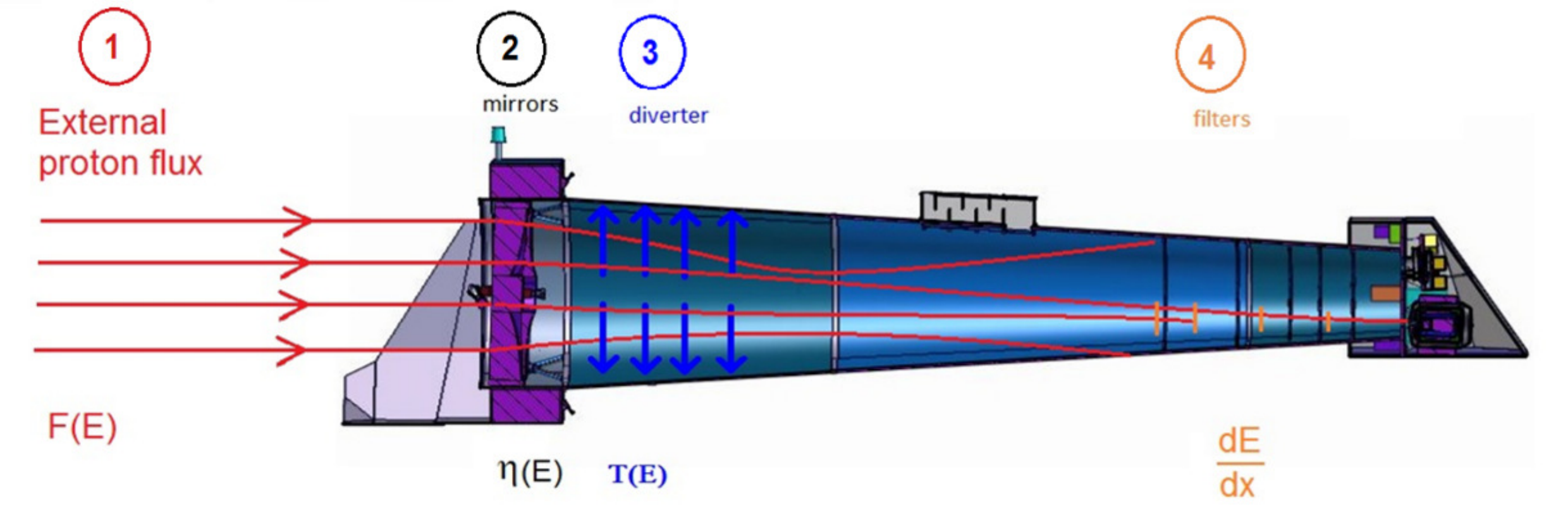}
\caption{Schematics of the steps followed by the soft protons through the telescope: 1) we start with the external soft protons environment, 2) the protons interact with the mirrors, with funnelling efficiency $\eta$, 
3) the protons are deflected by the magnetic diverter with transmission efficiency T, 4) the protons experience energy loss $\frac{dE}{dx}$ inside the thermal filters before reaching the detector.}
\label{fig:1} 
\end{figure*}

Here we estimate the background induced by these low energy particles exploiting a modular approach. We divide the problem into steps, according to the 
path followed by the protons towards the instrument (see Fig.~\ref{fig:1}): we first determine the external fluxes impacting on the optics (Sect.~\ref{sec:11}), then use ray-tracing simulations
to derive the mirrors funnelling efficiency (Sect.~\ref{sec:12}), then compute the energy lost crossing the thermal filters (Sect.~\ref{sec:13}), and finally put everything 
together to compute the fluxes expected on the instrument (Sect.~\ref{sec:14}).

In the whole treatment we neglect the presence of a magnetic diverter, which is actually foreseen among the mission requirements, since its characteristics and design (and thus its efficiency) have not been decided yet. 
As a consequence the fluxes obtained are representative of the ones expected in case no countermeasure is adopted to reduce the soft protons flux, and represents an overestimate of the real ones
that will be experienced by the \xifu. We use these flux values to derive an approximate requirement on the maximum energy of the particles that the diverter will have to be able to deflect.

\subsection{External fluxes}
\label{sec:11}
\begin{figure}[!tbp]

  \centering
  \begin{minipage}{0.45\textwidth}
    \includegraphics[width=\textwidth]{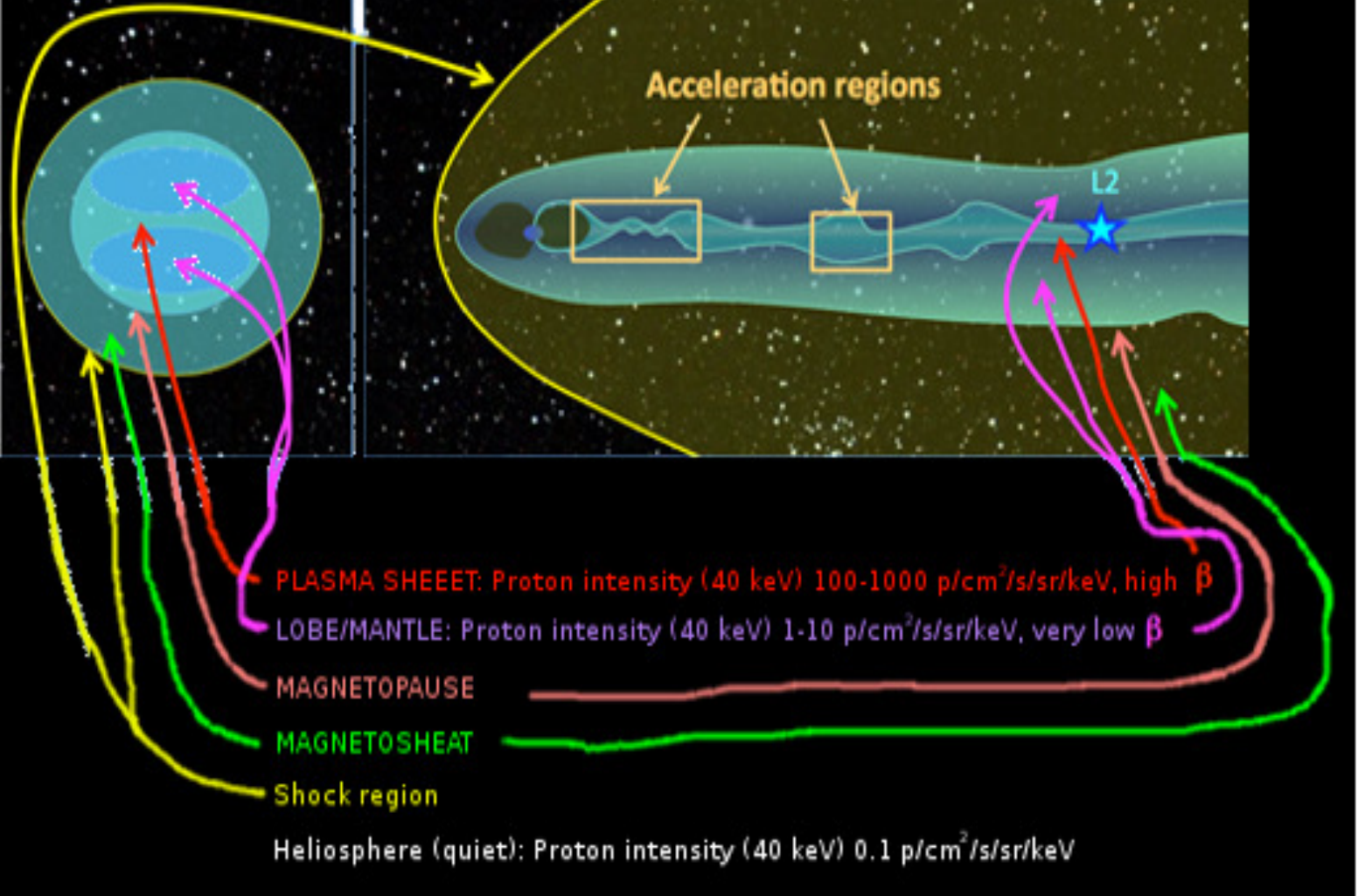}
  \end{minipage}
  \hfill
  \begin{minipage}{0.5\textwidth}
    \includegraphics[width=\textwidth]{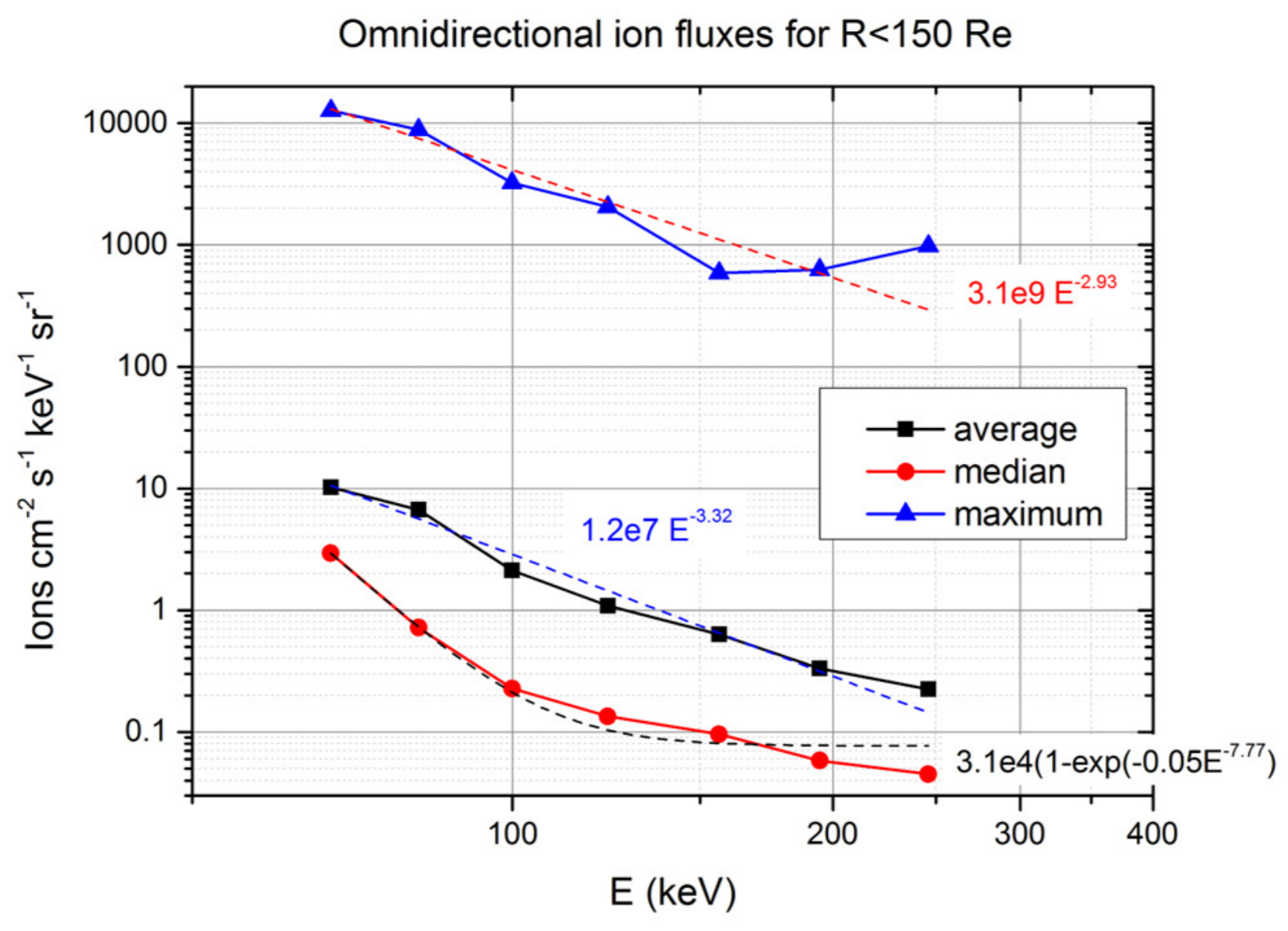}
  \end{minipage}
    \caption{\label{fig:2} Left - Schematic representation of the large scale magnetospheric structures/regions and their associated plasma regimes. 
    The $\beta$ plasma parameter is the ratio of the plasma thermal pressure over the magnetic pressure. Right - The average, median and maximum spectra measured by \geotail~beyond 150 $R_{e}$.}
  \end{figure}

The low energy environment in L2 is currently poorly known, complex, and highly dynamical. The Earth magnetotail in fact is constantly moving according to the solar wind and the magnetic field
influence, and the L2 point is swept by the moving tail. Furthermore, the different zones inside the tail are constantly changing shape and size (see Fig.~\ref{fig:2} - left). As a consequence, it is hard to predict the fluxes 
that will be experienced by a mission orbiting at L2. A part of the ESA tender AREMBES is dedicated specifically to this issue: here we exploit some preliminary data coming from the analysis of the \geotail~satellite
and some rough assumptions to produce two different estimates.

The first estimate assumes the spectral shape measured in the heliosphere, since it will be representative of the time periods \athena will be outside the magnetotail, $F _{Hel}\propto E^{-1.5}$ \cite{fisk2008}, 
and a flux at 80 keV averaged among the different magnetotail zones assuming that a fraction of time $>90\%$ is spent outside the plasma sheet, $F_{80~keV}\sim 10.5$ \partflux \cite{arembes}. 
It is worth to notice that, given the short separation between L1 and L2, this heliospheric flux is also representative of the L1 environment.

The second estimate has been produced using all the measurements obtained by the EPIC instrument onboard of \geotail~beyond 150 $R_{e}$. The orbit is different from the \athena one, but these data are the closest
to the L2 point available at present. The average flux measured at 80 keV is in accordance with our previous
assumption:​ $F_{80~keV}\sim 10$ \partflux. The spectral slope for the average case however is way steeper, $F_{Mag} \propto E^{-3.3}$ (see Fig.~\ref{fig:2} - right).
Note that the measured fluxes are omni-directional and that the lowest energy channel of the \geotail~EPIC instrument is sensitive to energies higher than 80 keV, so we were forced to extrapolate down to 1 keV the spectrum measured at higher energies.

\subsection{Mirrors funnelling efficiency}
\label{sec:12}

Protons impacting the X-ray mirrors at grazing incidence angles are known to be reflected in a process similar to the one experienced by photons. The scattered protons can not escape 
the telescope tube, and are funnelled towards the focal plane. We have performed two independent estimates for the funnelling efficiency of the optics 
defined as the ratio between the number of particles impacting on the detector per unit area $n_{det}$ and  the intensity of the proton flux at the optics $I_{inc}$ in $p~cm^{-2}~s^{-1}~sr^{-1}$.

\begin{itemize}

\item The first estimate is based on ray-tracing simulations for protons impacting on a model of the \athena optics, assuming a reflection efficiency and scattering obtained in elastic approximation and independent on the particle energy.   Incident directions are uniformly generated in a solid angle with a maximum off-axis of 20$^\circ$ and events reaching the focal plane are considered if they are contained in a 10 cm radius circle.  Under the first order assumption that the distribution at the focal plane is uniform we derive: 
\begin{equation}
\frac{n_{det}}{I_{inc}} \simeq 6.3 \times 10^{-3}~sr
\end{equation}

\item The second estimate is quite conservative, and assumes that the mirrors focalize protons with the same efficiency as 1 keV X-ray photons.  In this case, the funnelling efficiency can be easily computed from  the ratio between the intensity of CXB photons impacting on the mirrors $I_{ext}^{CXB}$ and the photon flux measured on the detector $f_{xifu}^{CXB}$: 
 \begin{equation}
\frac{n_{det}}{I_{inc}} \sim \frac{f_{xifu}^{CXB}}{I_{ext}^{CXB}} = \frac{\Omega(\theta) A_{opt}}{A_{det}}\sim 0.017~sr
\end{equation}
where where $\Omega(\theta)$ is the solid angle subtended by the optics (2$\times$10$^{-6}$ sr), $A_{opt}$ is the mirrors effective area at 1 keV, $A_{det}$ the detector geometrical area. The units for $f_{xifu}^{CXB}$ and $I_{ext}^{CXB}$  are $p~cm^{-2}~s^{-1}~keV^{-1}$ and $p~cm^{-2}~s^{-1}~keV^{-1}~sr^{-1}$, respectively.

\end{itemize}

\noindent The two estimates are in agreement with each other within a factor of 3, a result that given the complete independence of the two methods is quite remarkable. 
In the following we will use the first estimate, the second being just a first order check on the $\frac{n_{det}}{I_{inc}}$ value obtained with the simulations.

\subsection{Thermal filters}
\label{sec:13}
Soft protons aimed towards the detector will eventually reach the thermal filters in front of the detector, and lose part of their energy inside them. We are interested in particles that release energies inside the \xifu~
sensitivity band (0.2-12 keV). This means that particles with too high energies will be rejected by energy screening of the registered events, while protons below a certain energy will not  be able to pass through the filters. 
We expect then an intermediate energy range where protons contribute the most to this background component.

\begin{figure}[!tbp]

  \centering
  \begin{minipage}{0.49\textwidth}
    \includegraphics[width=\textwidth]{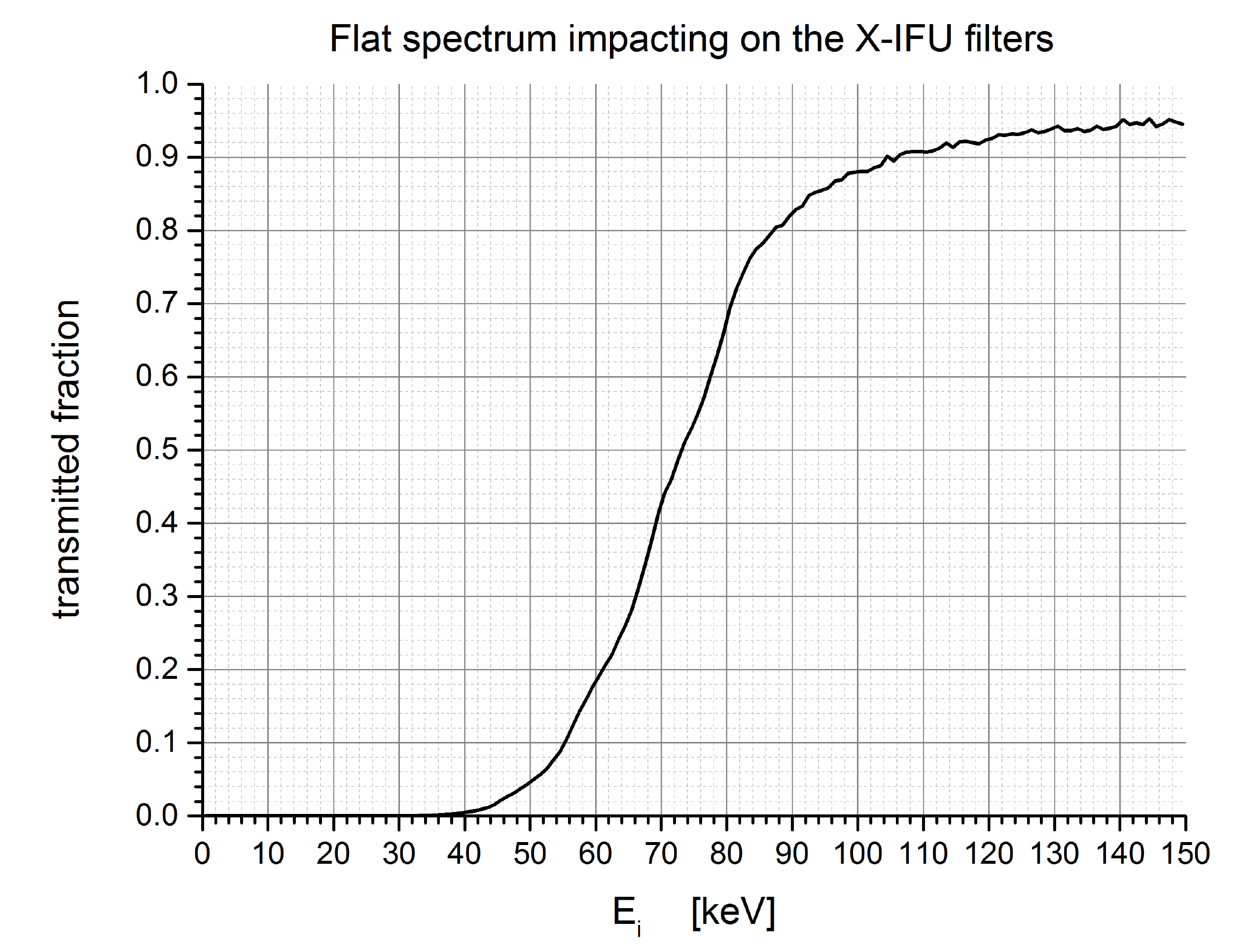}
  \end{minipage}
  \hfill
  \begin{minipage}{0.49\textwidth}
    \includegraphics[width=\textwidth]{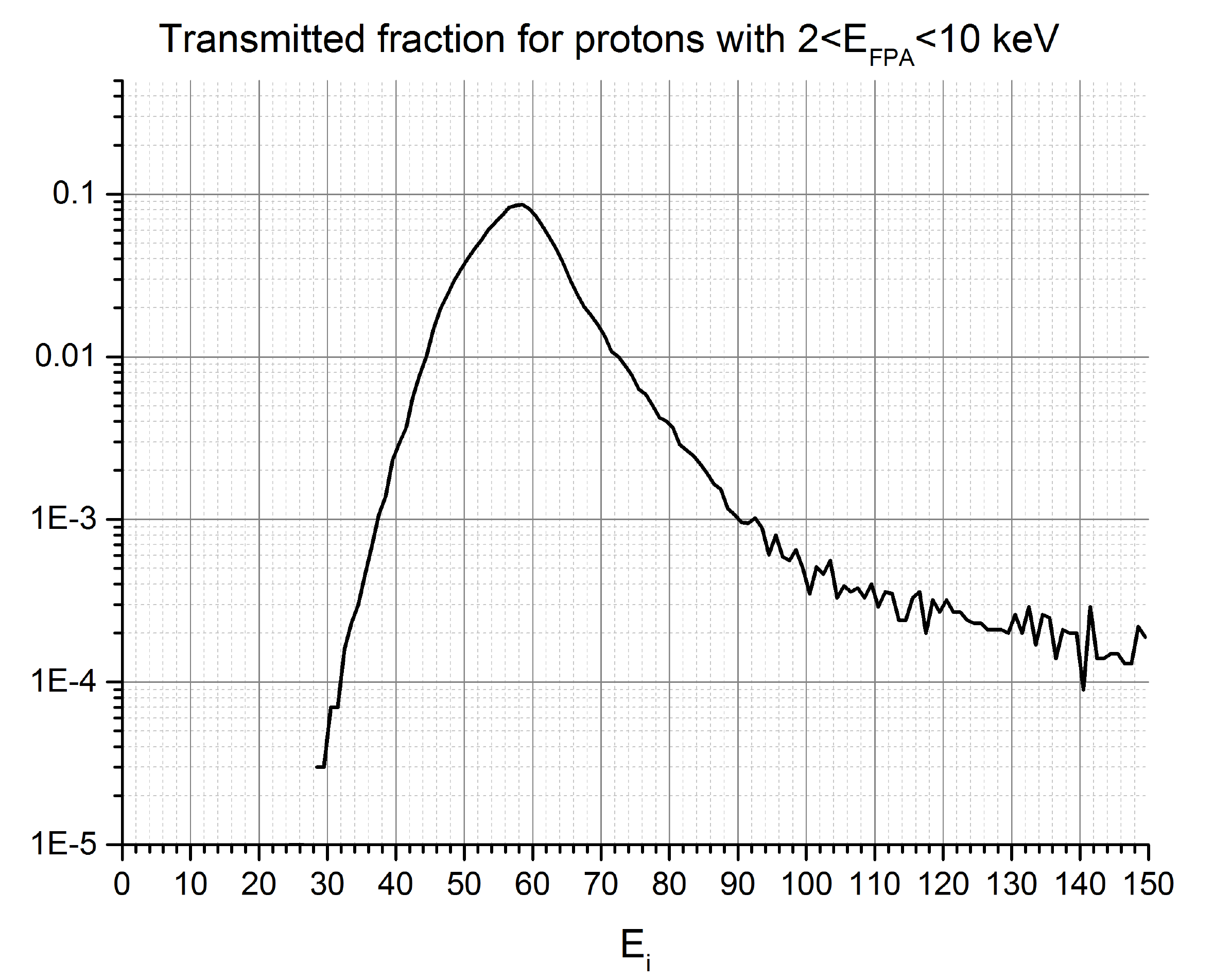}
  \end{minipage}
    \caption{\label{fig:3} Left - Transmission function for protons impacting on the X-IFU filters with a flat spectrum. 
    Right - Initial energy distribution of protons that reach the focal plane with energy $2<E_{FPA}<10~keV$.}
  \end{figure}

The exact transmission function of the filters has been calculated using Monte Carlo simulations. We reproduced inside a Geant4 simulation the current thermal filters baseline (a total of 0.28 $\mu m$ Kapton filter and 0.21 $\mu m$ Aluminum mesh, 
divided in 5 filters of identical thickness \cite{barbera}), and shoot a flat spectrum of protons between 1 and 150 keV onto these filters, towards the detector. We obtained then the transmission function of 
the \xifu~filters (Fig.~\ref{fig:3} - left), and the transmission function of the protons that reach the detector with energies $E_{FPA}$ inside the range indicated by the instrument background requirement (2-10 keV) (Fig.~\ref{fig:3} - right).

\subsection{The expected flux of soft protons on \xifu}
\label{sec:14}

\begin{figure*}
\centering
 \includegraphics[width=0.7\textwidth]{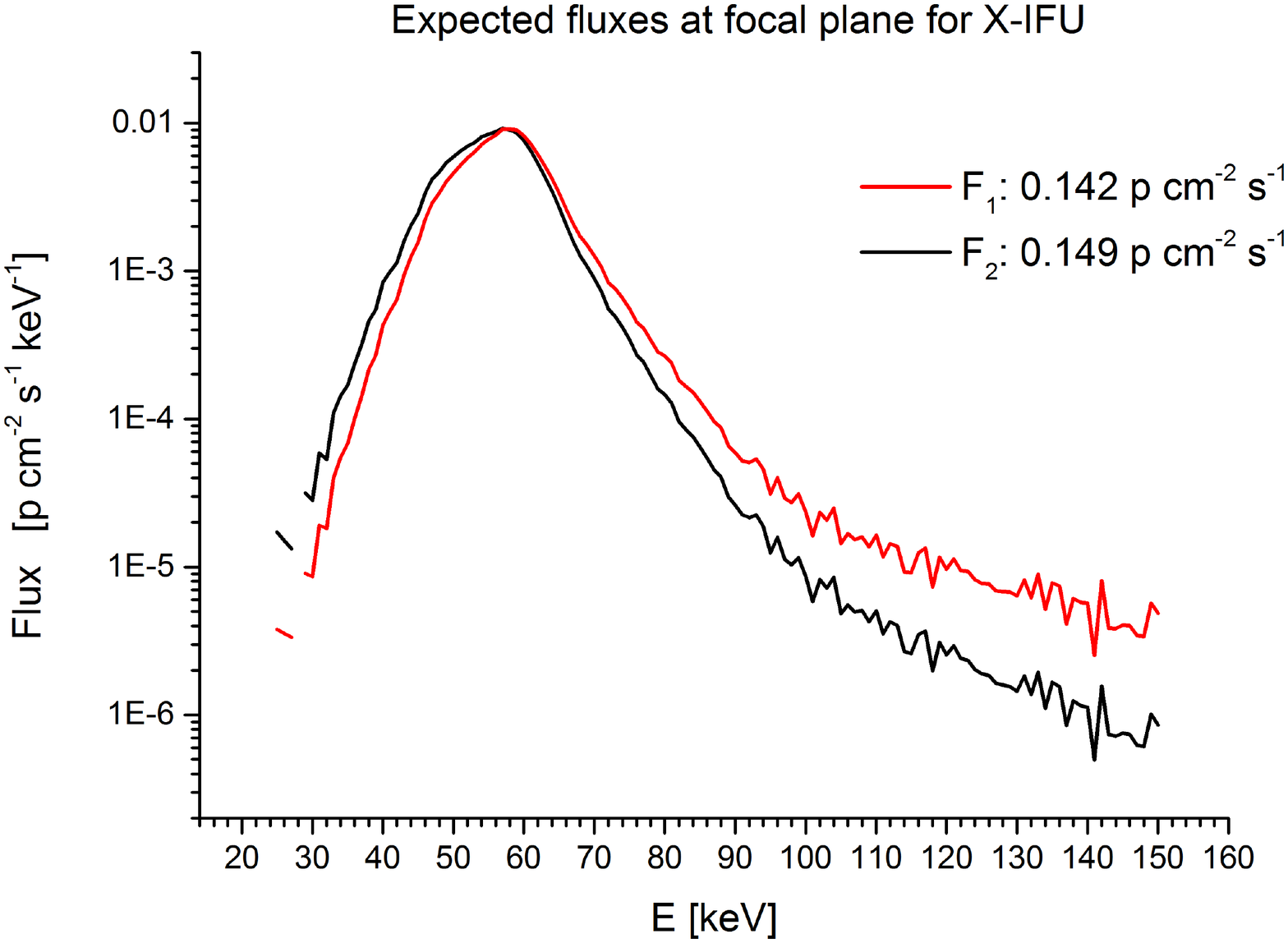}
\caption{Distribution of initial energies of protons that reach the focal plane with energy inside the X-IFU sensitivity band for the two external flux hypothesis described in the text.}
\label{fig:4} 
\end{figure*}

\begin{figure}[!tbp]

  \centering
  \begin{minipage}{\textwidth}
    \includegraphics[width=\textwidth]{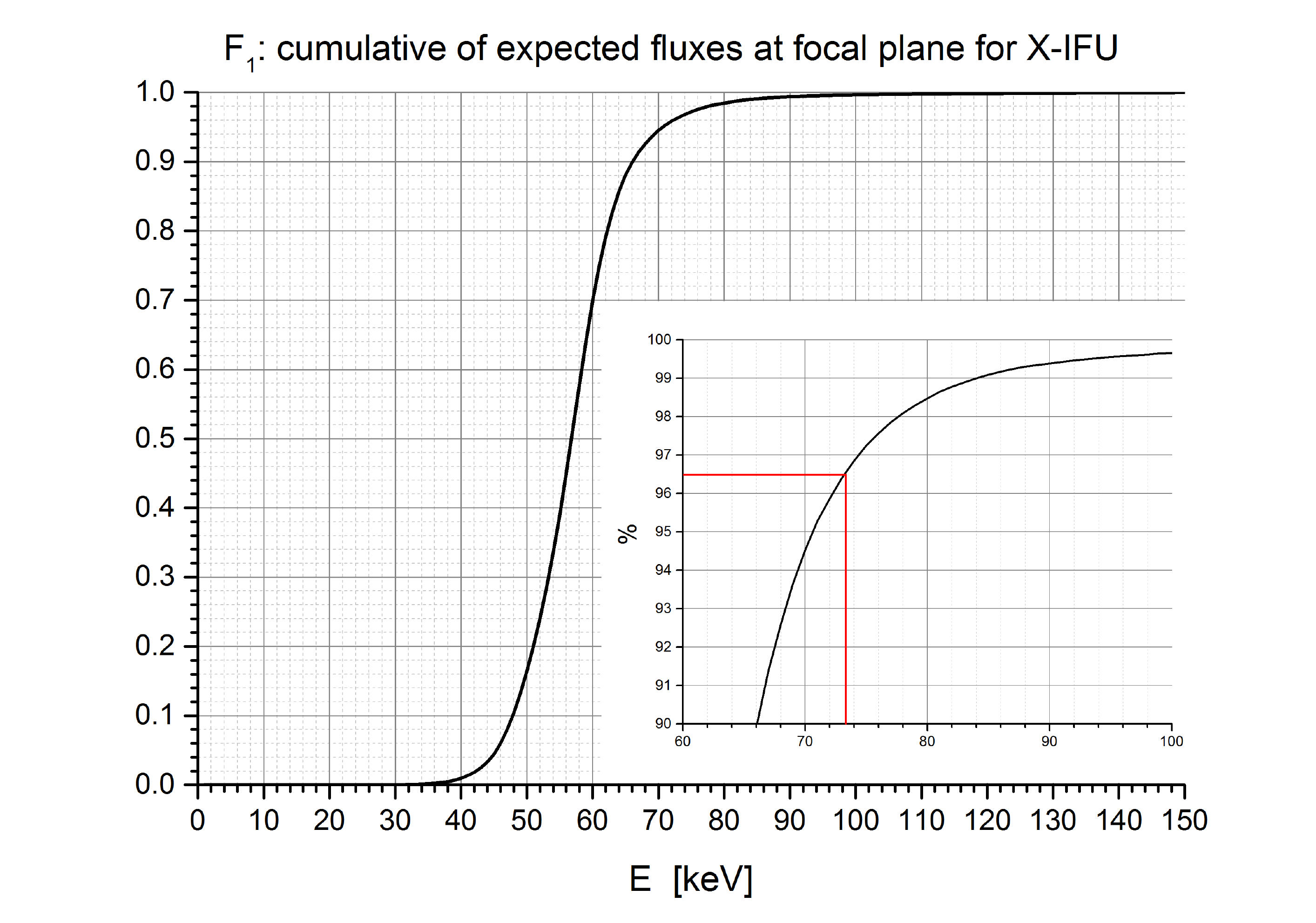}
  \end{minipage}
  \hfill
  \begin{minipage}{\textwidth}
    \includegraphics[width=\textwidth]{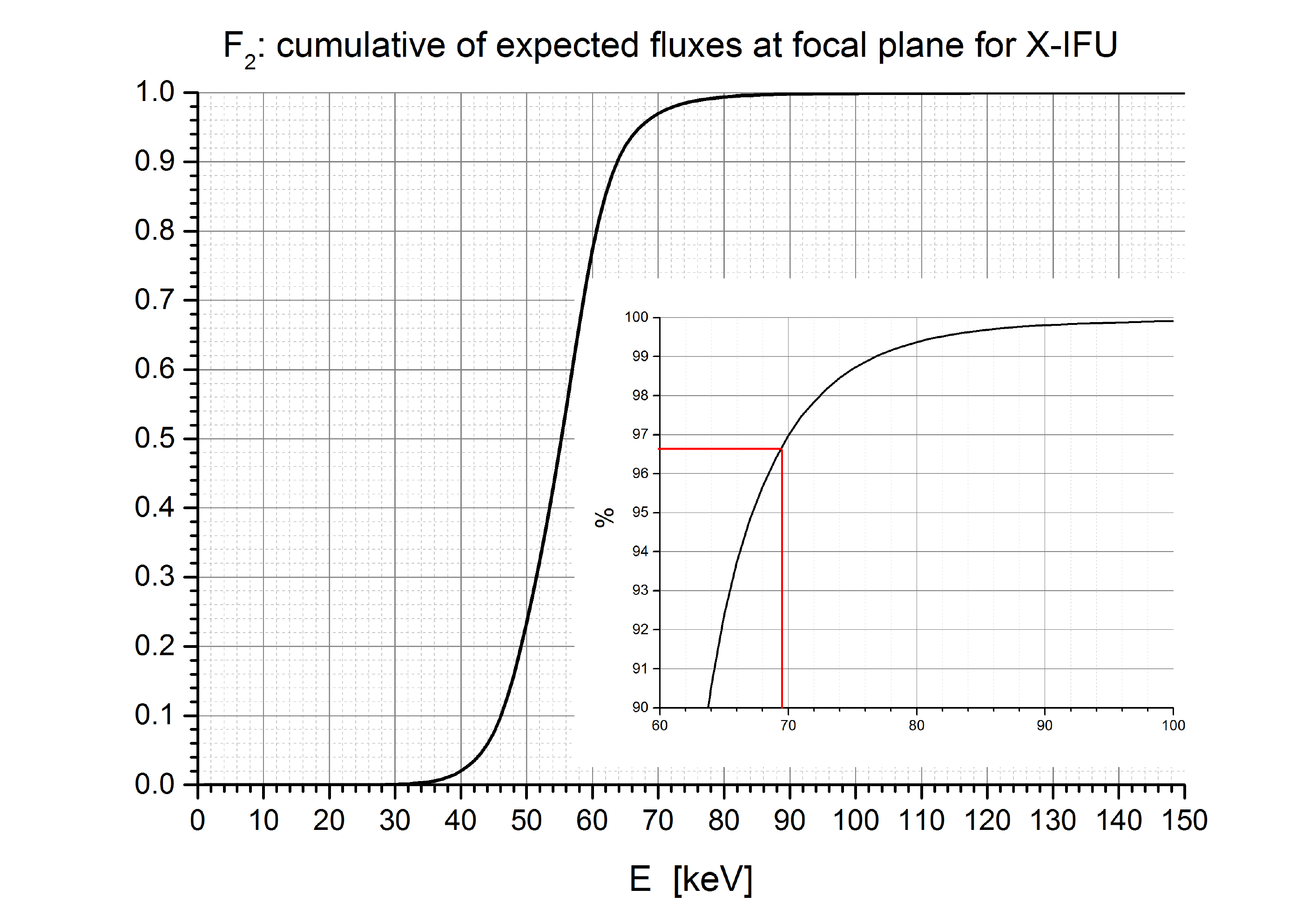}
  \end{minipage}
    \caption{\label{fig:5} Cumulative curves of the distributions shown in Fig.~\ref{fig:4} for the $F_{Hel}$ case (left), and the $F_{Mag}$ case (right). 
    The smaller plots are zoomings in the region where the approximate magnetic diverter requirement (the red line) is derived.}
  \end{figure}

Finally, multiplying the initial fluxes by the mirrors funnelling efficiency and the filters transmission function, we find the expected flux of background-inducing particles expected on the \xifu:
assuming $F_{Hel}$ (see Sect.~\ref{sec:11}) we obtain a total flux of $\sim 0.149~p~cm^{-2}~s^{-1}$, while with $F_{Mag}$ we find $\sim 0.142~p~cm^{-2}~s^{-1}$. In Fig.~\ref{fig:4} you can see the spectra
impacting on the detector in the two cases considered, while in Fig.~\ref{fig:5} the cumulative distributions of such fluxes for the $F_{Hel}$ and the $F_{Mag}$ cases are displayed. 

We want these fluxes to be reduced by the magnetic diverter below the requirement for soft protons induced background, $5\times10^{-3}~p~cm^{-2}~s^{-1}$ (which is 10\% of the NXB requirements), so we need to reach a rejection
 efficiency above 96.5\% for both cases. Looking at the cumulative distributions shown in Fig.~\ref{fig:5} we can see that these thresholds correspond to a magnetic diverter able to deflect every 
 particle with energy below $\sim 74~keV$ ($F_{Hel}$) and $\sim 70~keV$ ($F_{Mag}$).

\section{The Non X-ray Background}
\label{sec:2}

Particles with sufficient energy, typically cosmic rays with $E>\sim 100~MeV$, can cross the spacecraft and reach the focal plane from every direction. They deposit a fraction of their energy inside the absorbers,
creating background events that are hard to distinguish from photon-induced ones. 
In fact for a $\sim GeV$ proton typical energy losses are $\sim 1.16~keV/\mu m$ in Bismuth and $\sim 2.32~keV/\mu m$ in Gold, and for the \xifu~absorbers ($4~\mu m$ Bi, $1~\mu m$ Au) the most probable energy 
deposition lies inside the instrument energy band. Furthermore, the energy released while they cross the materials surrounding the detector goes in the creation 
of secondary particles which can in turn induce additional background \cite{lotti2012,lotti2014}. 
 
This NXB background has never been measured in L2 for X-ray detectors, so to estimate it we use Geant4 Monte Carlo simulations \cite{geant1,geant2,geant3}. In order to create a reliable Geant4 simulation 
the user has to specify several building blocks:

\begin{itemize}
\item An accurate mass model of the instrument and its surroundings
\item The model of the particle environment in which the mass model is expected to be placed
\item The physical models that reproduce the particles interactions with the mass model
\item Appropriate simulation settings
\end{itemize}

\noindent and last but not least a reasonable analysis framework that mimics what the actual pipeline of the instrument will do.

\begin{figure}[!tbp]

  \centering
  \begin{minipage}{0.49\textwidth}
    \includegraphics[width=\textwidth]{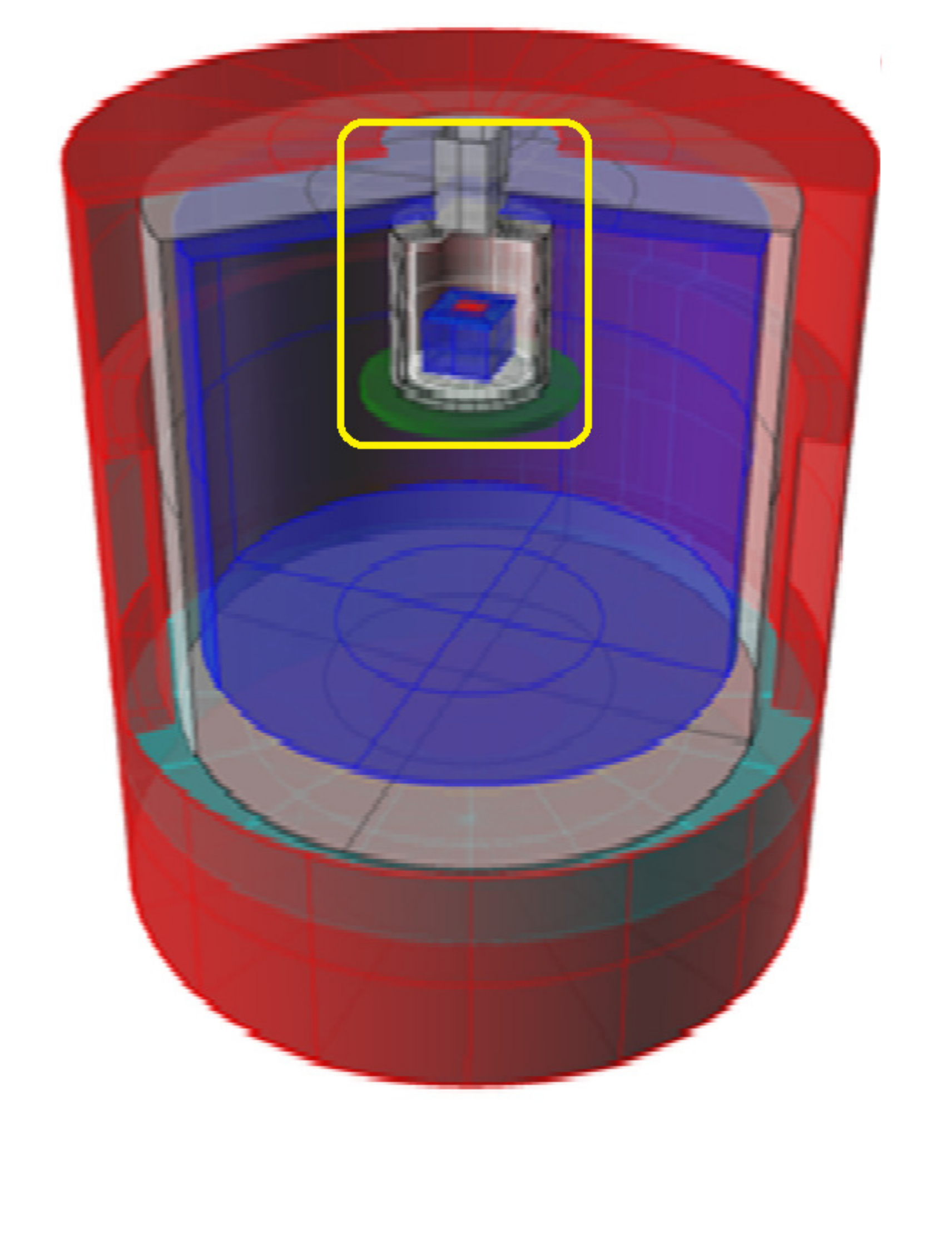}
  \end{minipage}
  \hfill
  \begin{minipage}{0.49\textwidth}
    \includegraphics[width=\textwidth]{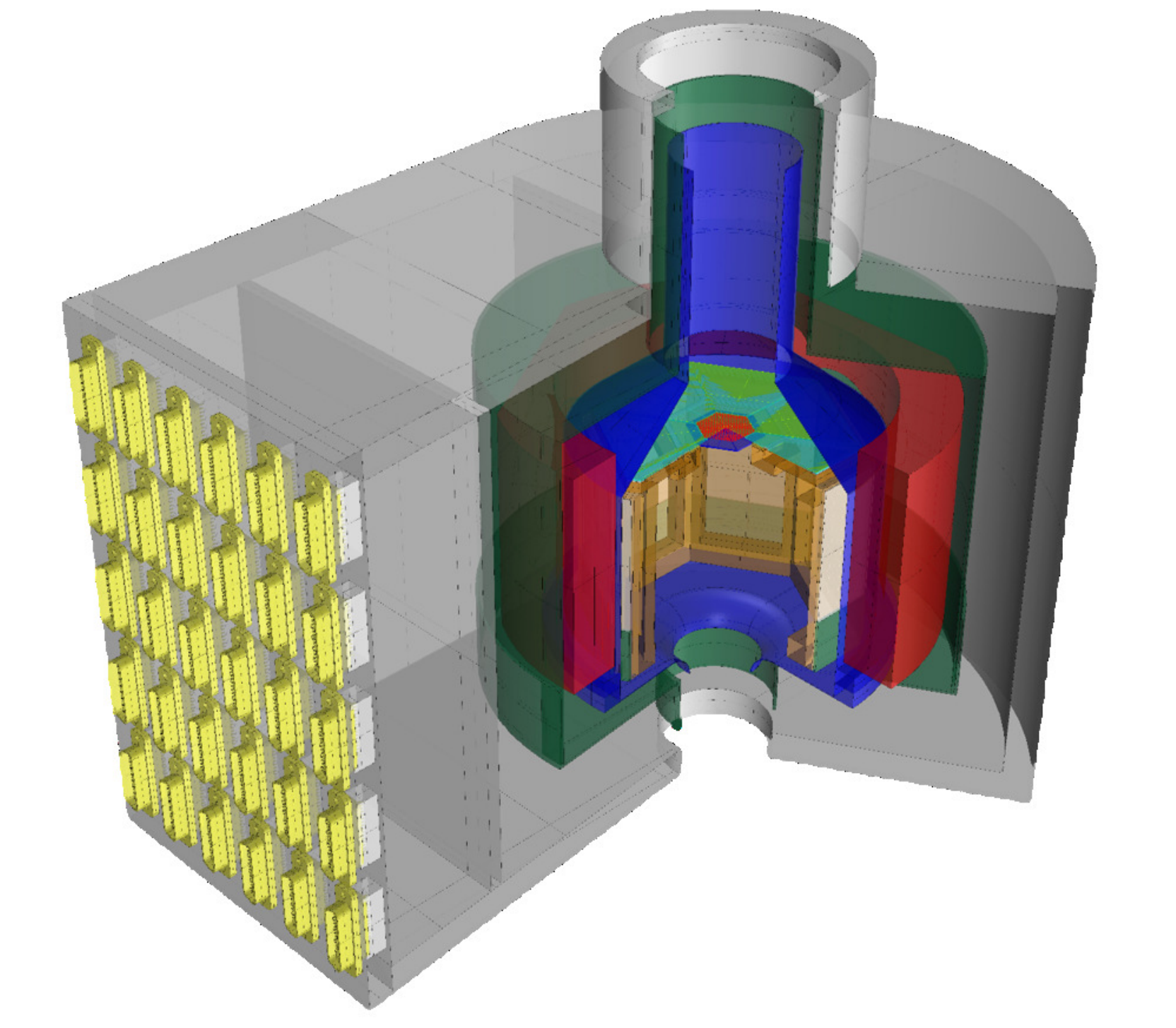}
  \end{minipage}
    \caption{\label{fig:6} Left - The cryostat mass model used in older Geant4 simulations. The FPA model is inside the yellow box. Right - The updated FPA mass model. It can be seen the much higher degree of complexity with respect to the previous one.}
  \end{figure}

An accurate mass model of the instrument, especially in the detector proximity, is crucial since the background level depends on the materials, their placement, their shapes, and on the total mass shielding the detector from radiation. The mass model
of the Focal Plane Assembly has been recently updated with respect to the simplified model used in the previous estimates. This update caused an increase in the mass surrounding the \xifu, which in turn generated an increased flux of secondary particles
towards the detector, and an increase in the unrejected background with respect to the previous estimates.
A comparison of the two mass models can be seen in Fig.~\ref{fig:6}.

\begin{figure*}
\centering
 \includegraphics[width=0.7\textwidth]{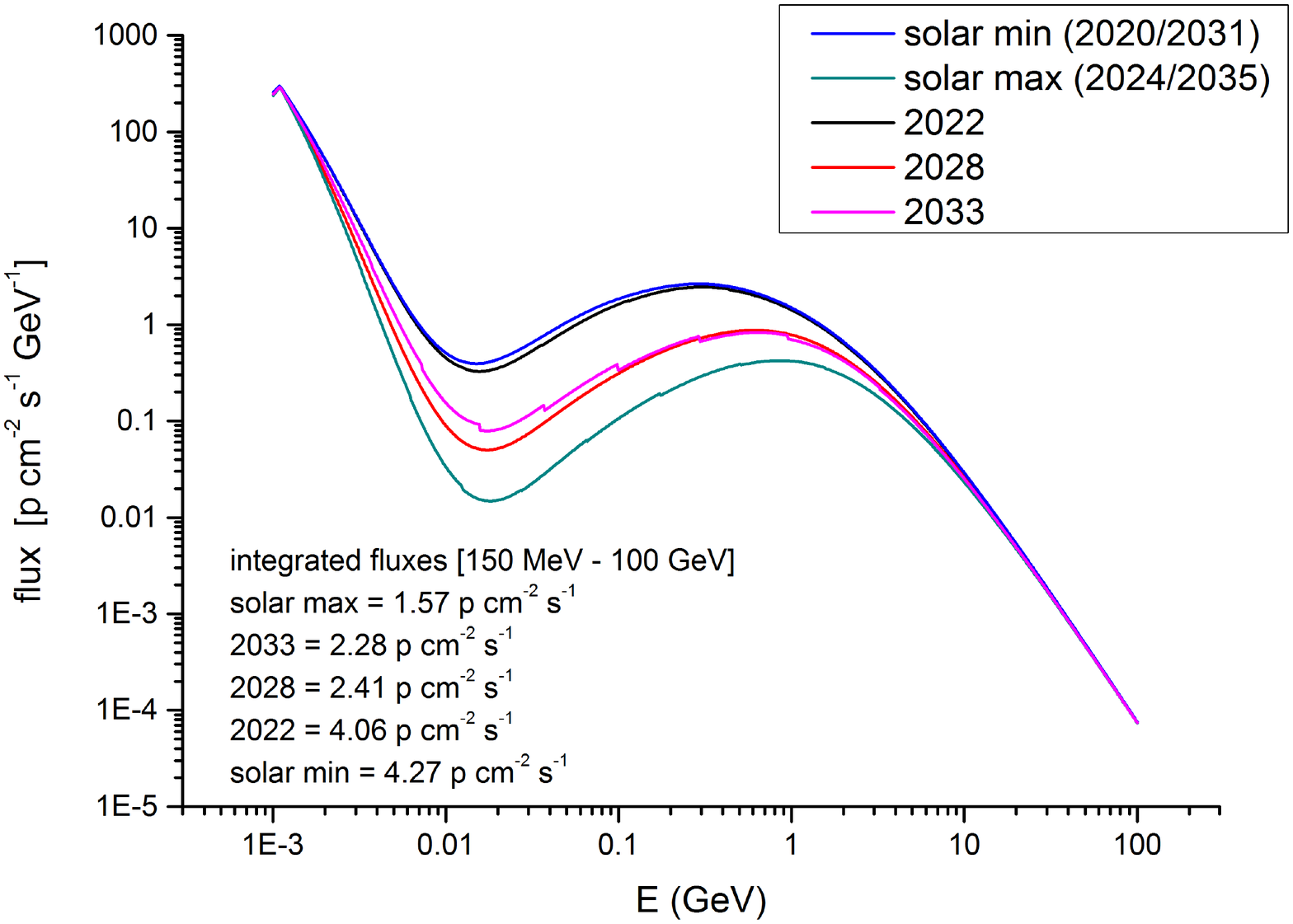}
\caption{Protons fluxes expected in L2 for several epochs, as reported by CREME96.}
\label{fig:7} 
\end{figure*}

The L2 environment is modeled using the cosmic rays protons spectrum reported by Creme96 \cite{creme1,creme2,creme3} for the solar minimum in interplanetary space, and is shown in Fig.~\ref{fig:7}. Since this spectrum corresponds to the maximum expected flux for the mission lifetime this represents a conservative estimate.

The simulations have been performed with Geant4 version 10.1, and have been tuned to have the highest accuracy from in the inner ones (i.e., 
the solids directly seen by the detector), which decreases into the outer zones. This is to ensure that the regions generating the biggest fraction of the unrejected
background are treated with the greatest care, while the interactions in the remaining mass model does not slow down the simulation too much.

The analysis framework handles the correct reconstruction of the events inside the detector pixel grid, the rejection of the background events based on event grading, and on the detection of events
in both the main detector and the CryoAC within given energy and time thresholds (anticoincidence).

\subsection{Background in the baseline configuration}
\label{sec:21}

Without applying any kind of background reduction technique the \xifu~would be subject to an unrejected particle flux of $0.56~p~cm^{-2}~s^{-1}~keV^{-1}$ in the 2-10 keV energy range (see Table~\ref{tab:1}). This value is 2 orders
of magnitude above the scientific requirement ($5\times10^{-3}~p~cm^{-2}~s^{-1}~keV^{-1}$ in the 2-10 keV energy band), so a Cryogenic AntiCoincidence detector is foreseen in the baseline configuration. This CryoAC is composed of four
TES detectors, with four large area absorbers (500 $\mu m$ thick), placed $<1~mm$ below the main detector, that allow to reject the majority of the high energy particles inducing this background value. 
With the insertion of this device is possible to reduce the unrejected background by $\sim 33$ times, reaching $1.7\times10^{-2}~p~cm^{-2}~s^{-1}~keV^{-1}$.

This residual background is composed mainly of secondary electrons that backscatter on the detector surface depositing a small fraction of their energy. The CryoAC can not possibly intercept such particles, thus 
we intervene on the materials surrounding the \xifu~to further reduce the unrejected background value, with passive shieldings that have a low secondary electrons yield and at the same time allow to block the electrons flux coming from the FPA.

\begin{figure}[!tbp]

  \centering
  \begin{minipage}{0.49\textwidth}
    \includegraphics[width=\textwidth]{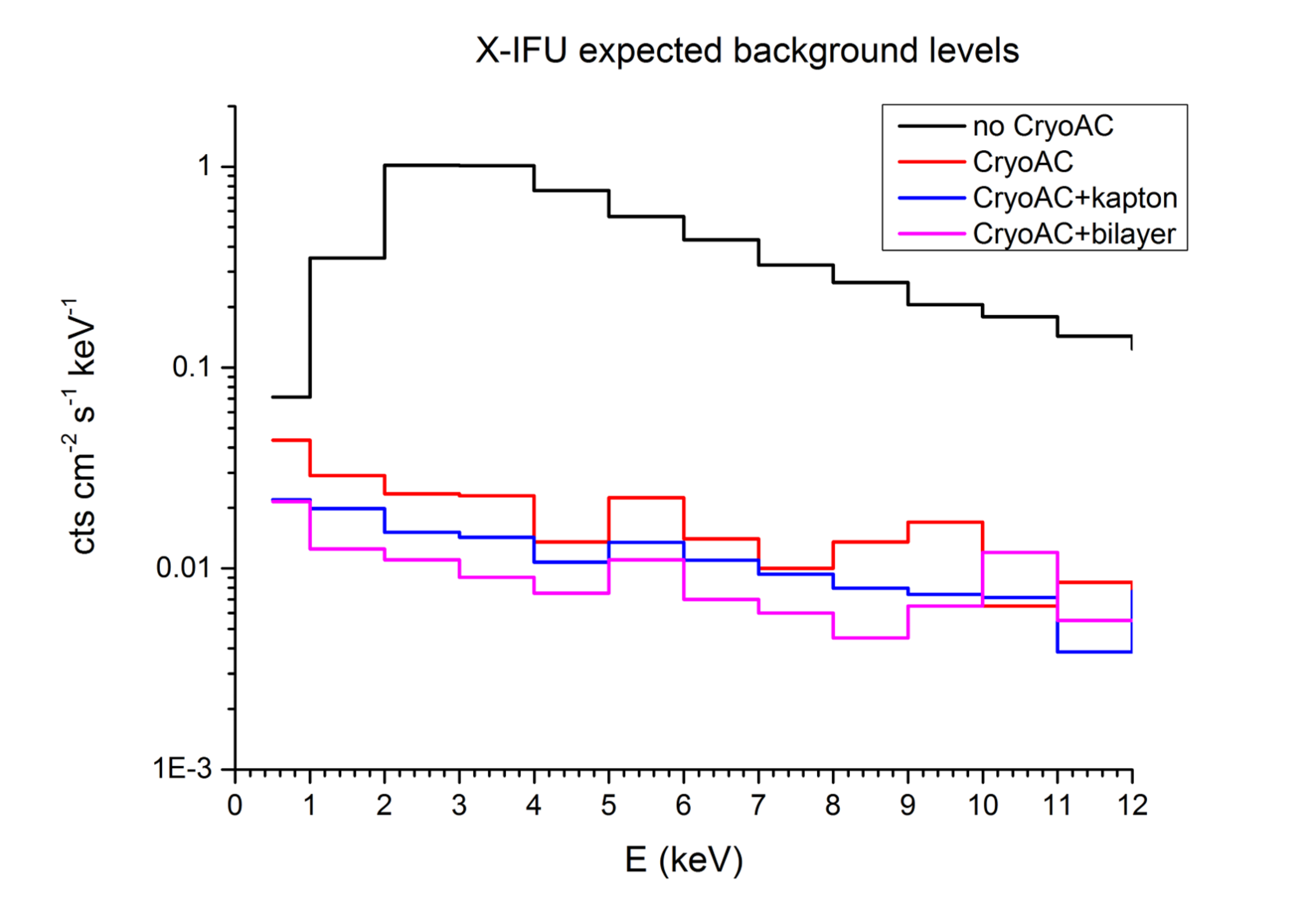}
  \end{minipage}
  \hfill
  \begin{minipage}{0.49\textwidth}
    \includegraphics[width=\textwidth]{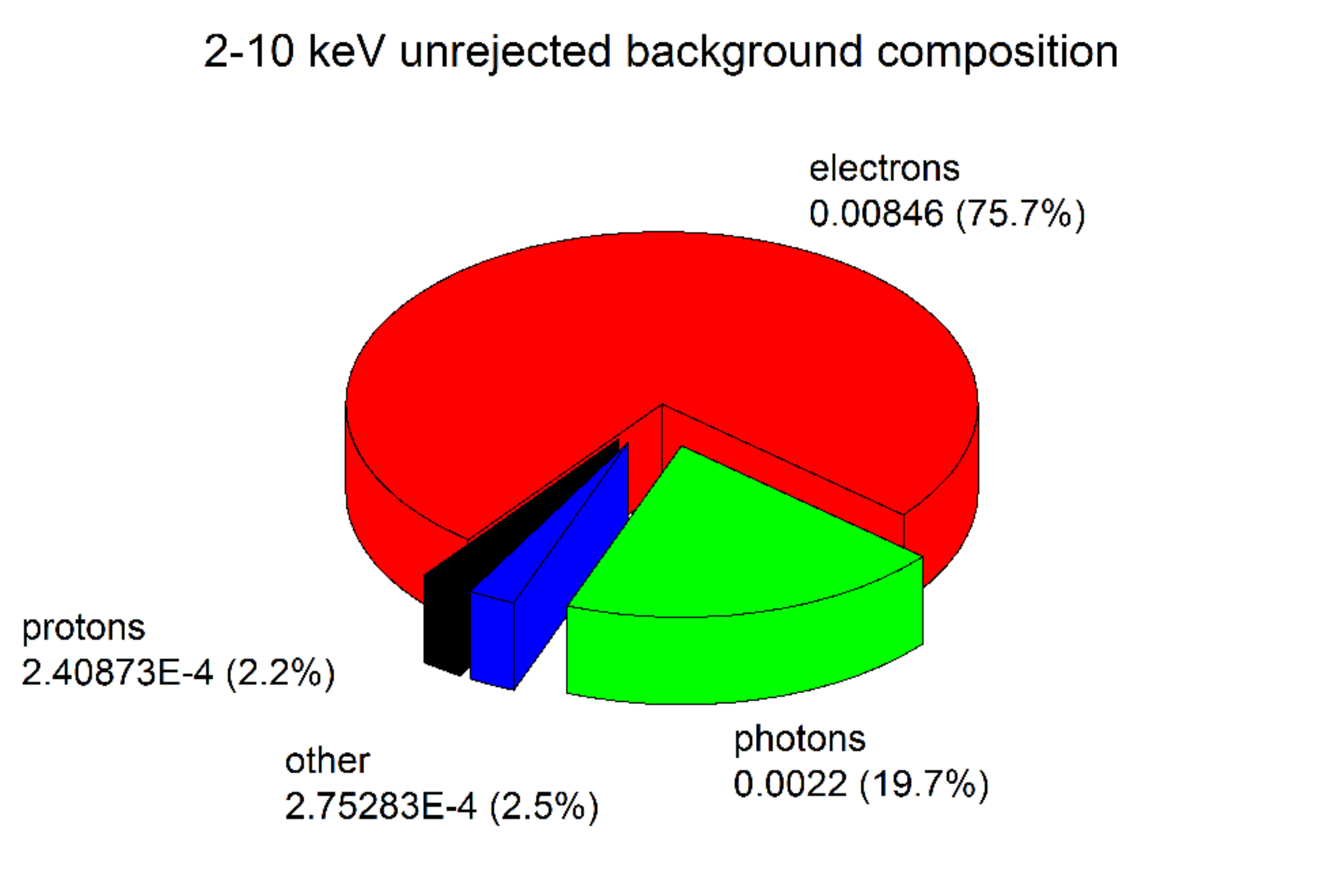}
  \end{minipage}
    \caption{\label{fig:8} Left - Unrejected background spectra without the CryoAC (black line), after the insertion of the CryoAC (red line), with the CryoAC and the Kapton shield (blue line), 
    and witht the improved passive shield described in Sect.~\ref{sec:22} (magenta line). Right - The composition of the background with the CryoAC and the Kapton shield (the blue line in Fig.~\ref{fig:8} - left).}
  \end{figure}

The baseline configuration foresees a Niobium shield (the blue solid in Fig.~\ref{fig:6} - right) aimed to shield the detector from magnetic fields that could degrade its energy resolution. 
Unfortunately Niobium is a high producer of secondary electrons, so a passive shield made of Kapton was introduced to reduce the flux of secondary electrons towards the detector. 
The background level in this configuration in the 2-10 keV energy band is $1.1\times10^{-2}~p~cm^{-2}~s^{-1}~keV^{-1}$. This background is induced mainly by secondary electrons 
($\sim75\%$) and photons ($\sim20\%$, half of which in the form of fluorescence lines from the niobium shield), as it can be seen in Fig.~\ref{fig:8} - right. 

The majority of these electrons are high energy electrons ($<\sim MeV$) that do not cross the detector, but instead backscatter on its surface releasing a small fraction of their energy. 
As a consequence the CryoAC device cannot intercept them, and the only reasonable strategy we have is to damp their production as much as possible with passive shieldings. 

In Fig.~\ref{fig:8} - left we show the different background levels when the detector is left without an anticoincidence device (black line), with the insertion of the CryoAC (red line), with the Kapton shield (blue line) and with an improved version of such passive shield
that will be discussed in Sect.~\ref{sec:22} (magenta line). In Table~\ref{tab:1} the corresponding integrated background levels are reported.

\begin{table}
\centering
\caption{Average background levels in the 2-10 keV energy band. The errors due to the statistic of the simulations are below 10\%.}
\label{tab:1} 
\begin{tabular}{lll}
\hline\noalign{\smallskip}
Configuration 		& Unrejected background & \\
\noalign{\smallskip}\hline\noalign{\smallskip}
Without CryoAC 		&  $0.57$ 		& $p~cm^{-2}~s^{-1}~keV^{-1}$\\
With CryoAC 		&  $1.7\times10^{-2}$ 	& $p~cm^{-2}~s^{-1}~keV^{-1}$ \\
CryoAC + Kapton 	& $1.1\times10^{-2}$ 	& $p~cm^{-2}~s^{-1}~keV^{-1}$ \\
CryoAC + Kapton-Bi 	&  $7.8\times10^{-3}$ 	& $p~cm^{-2}~s^{-1}~keV^{-1}$ \\

\noalign{\smallskip}\hline
\end{tabular}
\end{table}

\subsection{Tests on different FPA configurations}
\label{sec:22}
In order to reduce the flux of secondary particles towards the \xifu~we run several simulations to optimize the shield thickness and materials. We first tested the influence of the Kapton shield thickness, 
and found that a too thin Kapton shield is unable to efficiently block electrons generated in the Niobium, while above a thickness of 250 $\mu m$ the shield mass increases with no appreciable benefits. 

Roughly half of the secondary photons component is generated by 16 and 18 keV lines produced inside the Nb shield: when these photons impact the detector they induce the emission of 10.8 keV and 13 keV 
fluorescence photons from the Bi layer of the absorber. The 10.8 keV and 13 keV photons escape, leaving inside the detector a fixed amount of energy in the form of escape peaks. The remaining contribution is given 
by low energy photons that are completely absorbed, and by high energy photons that Compton scatter in the detector, leaving a small fraction of their energy.

To block these 16 and 18 keV photons, we tested several configurations of double/tri-layered passive shieldings (see  Table~\ref{tab:2}), in which one or two layers of additional thin layers of high-Z material 
were introduced between Nb and Kapton to increase the shield stopping power.

\noindent In summary, we found that:
\begin{itemize}

\item A few $\mu m$ of Tungsten placed between the Nb and the Kapton efficiently suppress the Nb lines. Unfortunately the Tungsten produces further fluorescences inside the instrument energy band.
\item Inserting a further layer of Bi to block these W fluorescences (a Kapton-Bi-W tri-layer shield) we eliminate the W lines, but we get L fluorescences from Bi in turn;

	\begin{itemize}
	\item Substituting Bi with SiC or $Si_{3}N_{4}$ we get rid of all fluorescences
	\item Taking out Kapton, so that the last surface seen by the detector is $Si_{3}N_{4}$, we have no fluorescences above 2 keV and a low background level. The 1.72 keV line from Si can however be a hindrance for the observation of AGNs at redshift ~2-3.
	\end{itemize}
	
\item Other bi-layers tested: Kapton-Bi, Kapton-$Si_{3}N_{4}$, Kapton-SiC

	\begin{itemize}
	\item A remarkable result was obtained using a bilayer made of 250 $\mu m$ of Kapton and 1.3 mm of $Si_{3}N_{4}$, roughly halving the photon component and reducing the total background by ~25\%
	\item The Kapton-Bi bilayer also brought a ~20\% background reduction. Furthermore, half of the photon background is concentrated in the Bi line at 10.8 keV. 
	\item The Kapton-SiC solution brought results similar to the previous two, but with an escape peak at 5.7 keV.
	\end{itemize}

\end{itemize}

\noindent The best result was obtained with the Kapton-Bi bilayer, given that it reduced the electrons flux by $\sim25\%$, and that half of the photon-induced background is concentrated in the 10.8 keV line, 
near the edge of the sensitivity band of the instrument. This background level is shown in Fig.~\ref{fig:8} as the magenta line. This solution is remarkable also since we already know it is feasible to cool
down the Bi to cryogenic temperatures.

\begin{sidewaystable}
\vspace{300pt}
\captionof{table}{Average background levels in the 2-10 keV energy band. Numbers are in units of $[\times10^{-3}~p~cm^{-2}~s^{-1}~keV^{-1}]$, and the errors due to the statistic of the simulations are below 10\% for the total background level. The last line indicates if in the residual background spectrum there are fluorescence lines or escape peaks (E.P.).}
\begin{tabularx}{1\textheight}{LLLLLLLLLL}
\label{tab:2}

Background [2-10 keV]     & 250 $\mu m$ Kapton    & 250 $\mu m$ Kapton + 10 $\mu m$ W     &  \tiny250 $\mu m$ Kapton + 20 $\mu m$ Bi + 10 $\mu m$ W    & \tiny250 $\mu m$ Kapton + 250 $\mu m$ SiC + 10 $\mu m$ W   & \tiny250 $\mu m$ Kapton + 300 $\mu m$ $Si_{3}N_{4}$ + 10 $\mu m$ W     &  10 $\mu m$ W + 300 $\mu m$ $Si_{3}N_{4}$     &\tiny 250 $\mu m$ Kapton + 1.3 $mm$ $Si_{3}N_{4}$   & 250 $\mu m$ Kapton +  20 $\mu m$ Bi   & 250 $\mu m$ Kapton + 1 $mm$ SiC \\
    \hline
Total               & 11            & 9.8                   & 8.8                           & 8.8                           & 8.8                  	& 8.4                       & 7.7                       & 7.8                 & 8.4 \\
\hline
Photons             & 2.2         & 1.9                   & 1.3                           & 1.4                           & 1.5                        	& 1                         & 1.2                       & 1.3                   & 1.7 \\
\hline
Electrons           & 8.5         & 7.7                   & 6.8                           & 7                            & 6.7                         	& 6.8                       & 6.1                       & 6.1                   & 6.2 \\
\hline
Lines presence      & E.P.        & W:8.4 keV, 9.6 keV   & Bi:10.8 keV                    & No                            & No                          & Si:1.72 keV               & No                        & Bi:10.8 keV            & E.P. \\

  \end{tabularx}
\end{sidewaystable}

\subsection{Further activities}
\label{sec:23}
There are several activities foreseen in the \ahead framework, which will not be treated in this paper since they will be the subject of separate publications or are yet to be completed.

The first one concerns the introduction of an electron filter above the X-IFU to shield it from backscattering electrons. Preliminary results \cite{lottiSPIE} has shown that using Al filters there is no 
 significant reduction of the backscattered electrons component. We obtain a $\sim 20\%$ background reduction above 2 keV using 50 nm Au, or 3 $\mu m$ of BCB. However 
the X-ray transmission of such filters is too low to be considered for implementation. These results depend on the reliability of the backscattering process inside Geant4, which is 
currently under test inside the \arembes ESA contract, so they will have to be confirmed once the software settings that are able to reproduce the available experimental results will be identified.

The second additional activity regards the feasibility study to use the CryoAC as an independent X-ray detector, and is addressed in a separate paper \cite{dandrea}. 
Finally, in the future we will investigate the impact of the insertion of lateral walls for the CryoAC, aimed to increase the rejection efficiency of such a device and to shield the detector also from backscattered particles.

\section{Summary and conclusions}
\label{sec:3}
We have performed a first estimate of both the main components of the particle-induced background for the \xifu~instrument, namely the Soft Protons and the NXB. We wanted to provide a preliminary assessment of the
strength of the magnetic diverter required to reduce the soft protons flux on the \xifu~down to the required level (see Sect.~\ref{sec:1}), and to optimize the current passive shielding design with the aim to 
further reduce the unrejected background value on the instrument (see Sect.~\ref{sec:2}).

Regarding the soft protons component we adopted a modular approach, dividing the problem into the single steps that 
map different stages of the protons interaction through the satellite. In this process the greatest source of uncertainty is the external flux impacting on the mirrors. 
We adopted two different spectral shapes, corresponding to the one expected in the quiet heliosphere ($F_{Hel}\propto E^{-1.5}$) and to the one obtained by preliminary data analysis of the \geotail~dataset 
($F_{Mag}\propto E^{-3.3}$) performed in the \arembes framework, and obtained two similar values for the expected flux of particles depositing energy inside the sensitivity band on the \xifu: $0.14$ and $0.15~p~cm^{-2}~s^{-1}$
for the $F_{Hel}$ and the $F_{Mag}$ cases, respectively.
Both these estimates are above the requirement for the soft protons induced background ($5\times10^{-3}~p~cm^{-2}~s^{-1}$ in the 2-10 keV energy band), so we must adopt a magnetic diverter to deflect them away from the instrument FoV. 
The requirement for such a diverter is to be able to deflect $>96.5\%$ of the incoming flux for the both cases, 
corresponding to the deflection of particles up to 74 keV in the $F_{Hel}$ case, and up to 70 keV if we assume $F_{Mag}$. 

Regarding the NXB component, after the update of the FPA mass model and a finer tuning of the Geant4 settings, we estimate in the baseline configuration $1.1\times10^{-3}~p~cm^{-2}~s^{-1}~keV^{-1}$ 
in the 2-10 keV energy band, mainly induced by secondary electrons and photons. 
Starting from here we tested several alternative configurations of the passive shielding, and found that we can obtain an unrejected background level of $7.8\times10^{-3}~p~cm^{-2}~s^{-1}~keV^{-1}$
using a passive shield made of a thin layer (20 $\mu m$) of Bismuth and 250 $\mu m$ of Kapton.

It is to be remarked that this background estimates are referred to the expected maximum level of GCR flux (i.e., during the solar minimum), and it is thus representative of the worst conditions that \athena will experiment during its lifetime.

The estimates presented here are to be considered still preliminary, since there are several open points that will be addressed in the near future. 

Regarding the Soft Protons background the greatest uncertainty regards the external fluxes impacting on the mirrors (see Sect.~\ref{sec:11}), which will require the modeling of the 
magnetospheric structures at the L2 distances as a function of Earth rotation, 
geomagnetic activity and solar wind conditions and the characterization of the spectra of 50-100 keV ions accelerated in the magnetosphere with respect to the observation regions,
including the interplanetary medium. This activity is currently being conducted in the \arembes framework exploiting data from several satellites like \geotail, \wind, \artemis, \stereo~and \isee.

Besides that, the physical models describing the protons reflection on the mirror shells will have to be tested against experimental results and possibly improved, with a consequent 
update of the estimate of the mirrors funnelling efficiency. This activity is also part of the \arembes ESA contract, and of another recently issued ESA tender (\exacrad).

Finally the current estimates return an integrated flux level, while a redistribution matrix for the protons passing through \athena filters is under construction, which will allow 
to determine the spectrum of the soft protons impacting on the detectors. This will be the subject of a separate paper (\cite{dandrea}, in preparation) that will take into account
also different magnetic diverter configurations and efficiencies.

Regarding the NXB, the processes involved in the Geant4 simulations are under verification and validation with the aim to improve the agreement with experimental results and determine 
the optimal settings to run the simulations with. This activity is also part of the \arembes contract, together with the update of the remaining part of the mass model that became obsolete 
with the recent availability of new information on the \xifu~cryostat design.

\begin{acknowledgements}
Part of the research leading to these results has received funding from the European Union’s Horizon 2020 Programme under the \ahead project (grant agreement n. 654215). 
All the results described in this paper have been reported during the
\ahead background workshop, organized with the support of the EU Horizon 2020 Programme (grant agreement n. 654215).
\end{acknowledgements}



\end{document}